\documentclass[sn-aps]{sn-jnl}
\jyear{2022}
\usepackage{caption}
\usepackage{subcaption}
\begin{document}
\title[ ]{Understanding the direct detection of charged particles with SiPMs}

\author*[1]{\fnm{F.} \sur{Carnesecchi}}\email{francesca.carnesecchi@cern.ch}

\author*[2]{\fnm{G.} \sur{Vignola}}\email{gianpiero.vignola@desy.de}

\author[3]{\fnm{N.} \sur{Agrawal}}
\author[4]{\fnm{A.} \sur{Alici}}
\author[3]{\fnm{P.} \sur{Antonioli}}
\author[4]{\fnm{S.} \sur{Arcelli}}
\author[4]{\fnm{F.} \sur{Bellini}}
\author[3]{\fnm{D.} \sur{Cavazza}}
\author[4]{\fnm{L.} \sur{Cifarelli}}
\author[4]{\fnm{M.} \sur{Colocci}}
\author[5]{\fnm{S.} \sur{Durando}}
\author[4]{\fnm{F.} \sur{Ercolessi}}
\author[8]{\fnm{A.} \sur{Ficorella}}
\author[4]{\fnm{C.} \sur{Fraticelli}}
\author[6]{\fnm{M.} \sur{Garbini}}
\author[4]{\fnm{M.} \sur{Giacalone}}
\author[8]{\fnm{A.} \sur{Gola}}
\author[3]{\fnm{D.} \sur{Hatzifotiadou}}
\author[4]{\fnm{N.} \sur{Jacazio}}
\author[3]{\fnm{A.} \sur{Margotti}}
\author[4]{\fnm{G.} \sur{Malfattore}}
\author[3]{\fnm{R.} \sur{Nania}}
\author[3]{\fnm{F.} \sur{Noferini}}
\author[8]{\fnm{G.} \sur{Paternoster}}
\author[3]{\fnm{O.} \sur{Pinazza}}
\author[3]{\fnm{R.} \sur{Preghenella}}
\author[3]{\fnm{R.} \sur{Rath}}
\author[7]{\fnm{R.} \sur{Ricci}}
\author[3]{\fnm{L.} \sur{Rignanese}}
\author[4]{\fnm{N.} \sur{Rubini}}
\author[4]{\fnm{B.} \sur{Sabiu}}
\author[3]{\fnm{E.} \sur{Scapparone}}
\author[4]{\fnm{G.} \sur{Scioli}}
\author[4]{\fnm{S.} \sur{Strazzi}}
\author[3]{\fnm{S.} \sur{Tripathy}}
\author[4]{\fnm{A.} \sur{Zichichi}}

\affil[1]{\orgname{CERN}, \orgaddress{\street{Esplanade des Particules 1}, \city{Geneva}, \postcode{1211 Geneva 23},  \country{Switzerland}}}

\affil[2]{\orgname{Deutsches Elektronen-Synchrotron DESY}, \orgaddress{\street{Notkestraße 85}, \city{Hamburg}, \postcode{22607},  \country{Germany}}}

\affil[3]{\orgdiv{Sezione di Bologna}, \orgname{Istituto Nazionale di Fisica Nucleare}, \orgaddress{\street{viale Carlo Berti Pichat 6/2}, \city{Bologna}, \postcode{40127}, \country{Italy}}}

\affil[4]{\orgdiv{Dipartmento di Fisica e Astronomia "A. Righi"}, \orgname{University of Bologna}, \orgaddress{\street{viale Carlo Berti PIchat 6/2}, \city{Bologna}, \postcode{40127}, \country{Italy}}}

\affil[5]{\orgdiv{Dipartimento di elettronica e comunicazioni}, \orgname{Politecnico di Torino}, \orgaddress{\street{Corso Duca degli Abruzzi, 24}, \city{Torino}, \postcode{40127}, \country{Italy}}}

\affil[6]{ \orgname{Museo Storico della Fisica e Centro Studi Enrico Fermi}, \orgaddress{\street{Via Panisperna 89 A}, \city{Roma}, \postcode{10129}, \country{Italy}}}

\affil[7]{\orgdiv{Dipartmento di Fisica }, \orgname{University of Salerno}, \orgaddress{\street{Via Giovanni Paolo II, 132}, \city{Salerno}, \postcode{84084}, \country{Italy}}}

\affil[8]{\orgname{Fondazione Bruno Kessler}, \orgaddress{\street{Via Sommarive, 18}}, \orgaddress{\city{Povo}, \postcode{38123}, \country{Italy}}}


\abstract{In this paper evidence that the increased response of SiPM sensors to the passage of charged particles is related mainly to Cherenkov light produced in the protection layer is reported. The response and timing properties of sensors with different protection layers have been studied. }

\keywords{SiPM, tracking}

\maketitle

\section{Introduction}\label{sec1}

In a recent paper  \cite{SiPM1} the response of different Silicon PhotoMultipliers (SiPMs) to the passage of a charged particle was studied in detail \footnote{In \cite{SiPM1} other observations of the same phenomenon are also reported.}. The measurements highlighted two main results: i) an excess of number of pixels (SPADs) firing related to the passage of the particle\footnote{Excess expressed in Crosstalk, CT, measurements} with respect to the expectations of a single one (and the relative higher efficiency observed, w.r.t. a simple  geometry factor)  and ii) time resolutions down to 40 ps, depending on the device under test and improving with the number of fired SPADs.
The paper indicated few possible explanations for the observed excess of fired SPADs, such as different reaction processes in the bulk of the sensor structure or an effect of Cherenkov light due to the passage of the particle through the protection layer of the SiPM.

In this work, results of a new test beam are reported which demonstrate the presence of Cherenkov light as the main reason for the observed excess. A comparison of prototype SiPMs with different (in thickness and material) protection layers is also reported.

\section{Experimental setup}\label{sec2}

\subsection{Detectors} 
\label{sec:detectors}
For the present study available NUV-HD-RH SiPM produced by Fondazione Bruno Kessler (FBK) were used  \cite{2020Mazzi}.
These detectors are based on the NUV-HD technology \cite{2019Gola} and have an active area of 1$\times$1 mm$^2$, hexagonal pixel, with an equivalent rectangular pixel pitch of 20 $\mu \text{m}$, 
2444 number of SPADs,  72 $\%$ fill factor and
V$_{bd}$  33.0 $\pm$0.1 V \cite{SiPM1}.  The detectors were produced with three different protection layers of 1 and 1.5 mm silicon resin (refraction index 1.5, named SR1 and SR15)  and 1 mm in epoxy resin (refraction index 1.53, named ER1). Notice that the thickness refers to the support board and, since the sensor itself is 550 $\mu$m thick, the effective protection layer on top of the sensor is 450  and 950 $\mu$m for the 1 mm and 1.5 mm respectively.  A fourth prototype was produced without any protection resin (named WR). 

Each sensor is part of a structure of six nearby different SiPMs, all with the same protection layer. The SR1 sensor was already tested \cite{SiPM1}, but the protection layer was from a different resin producer.

\subsection{Beam test setup} 
\label{subsec:tb}
The SiPMs response has been studied with MIPs (Minimum Ioinizing Particles) at the T10 beamline of CERN-PS in July 2022. The beam was mainly composed of protons and $\pi^+$ with a momentum of 12 GeV/c. 

The telescope was made of four sensors: two SiPMs under test and two  LGAD detectors (1x1 mm$^2$ area and 35 $\mu m$ or 25 $\mu m$ thick prototypes) \cite{LGAD}; the latter are used as trigger for the beam particles and to evaluate the time resolution of the SiPM. 

The whole setup was enclosed in a dark box at room temperature. It is worth noticing that during the tests, the temperature ranged between 30 and 38 degrees which is higher compared to the characterization temperature usually employed for SiPMs. 

The SiPMs signals were independently amplified by XLEe with gain factor of about 40. The trigger was defined as the coincidence of the two LGADS in the telescope. At each trigger, all four waveforms were stored using a Lecroy Wave-Runner 9404M-MS digital oscilloscope\footnote{Lecroy WaveRunner datasheet:

\href{https://teledynelecroy.com/oscilloscope/waverunner-9000-oscilloscopes/waverunner-9404m-ms}{https://teledynelecroy.com/oscilloscope/waverunner-9000-oscilloscopes/waverunner-9404m-ms}}. 
For the final offline analysis, the oscilloscope bandwidth was set to 1 GHz, as for \cite{SiPM1}. 

\subsection{Signal selection}
\label{subsec:signal}

Signal and Dark Counts (DC) events selections proceed through different steps, similar to those described in \cite{SiPM1}. 
Given the LGADs trigger condition time (t$_0$), the signal events are those with a SiPM signal in a window of $\pm$ 2 ns from the trigger.
The DC events are defined as events in a region before the trigger (-25 ns to -5 ns): these are used to determine the possible contamination of noise events in the signal region. 
A further cut was applied on the SiPM signals by removing the few events with important residuals of previous signals (DC or MIP) in a time window of 8 ns before the signal zone (-10 ns to -2 ns from the trigger).

\begin{figure}[h!]
        \centering%
        \includegraphics [width=0.8\textwidth]{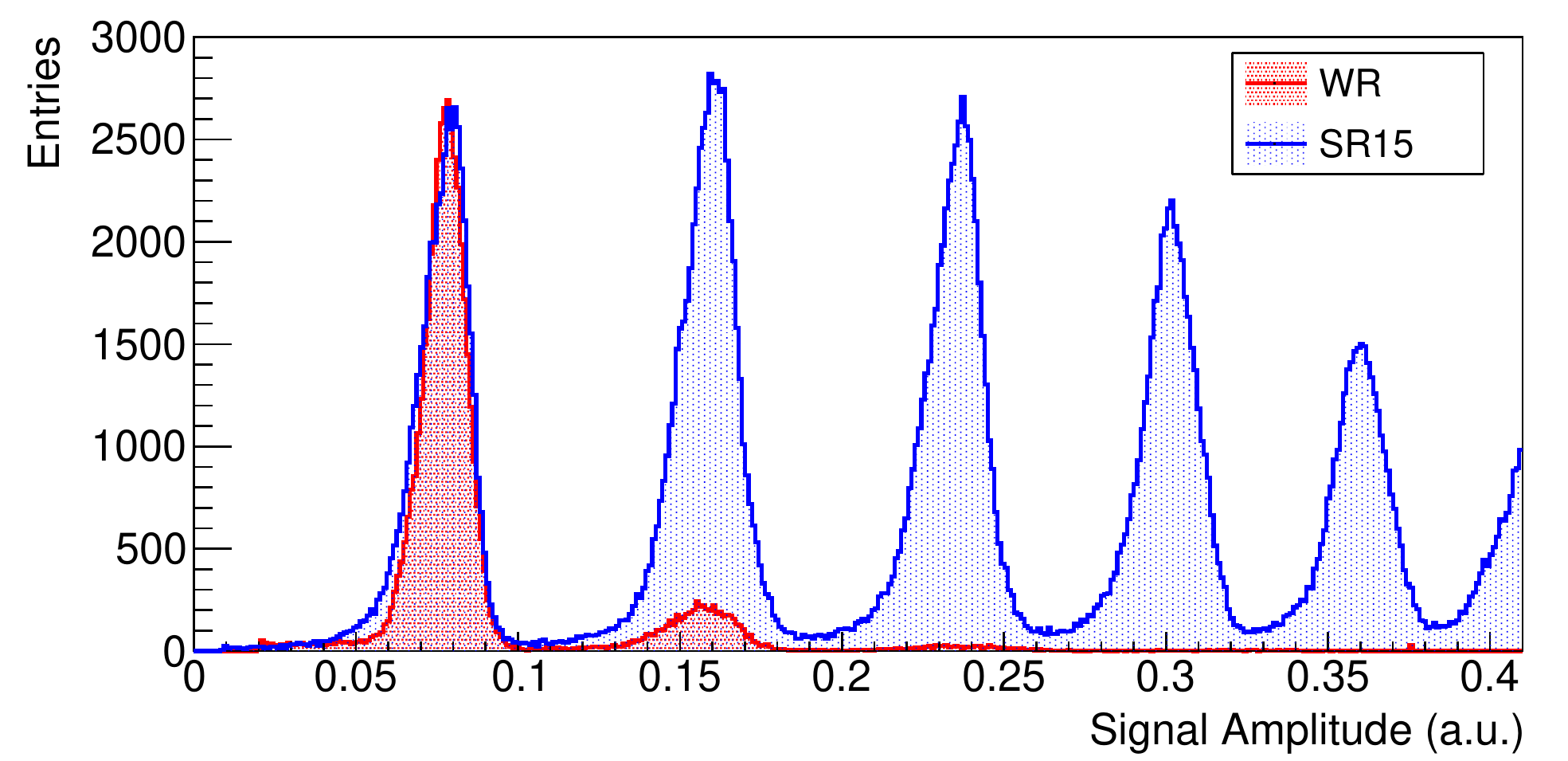}
        \caption{Example of MIPs Signals Amplitudes measured for the sensor Without Resin (WR, red) compared with the one with Silicon Resin and thickness of 1.5 mm (SR15, blue). In both cases the particles impinged the sensor from the front and an OV of $\sim$ 6 V was used. Histograms are normalised to the first peak.}
          \label{fig:signal}
\end{figure}

An example of signal amplitudes distribution after amplification and selection is shown for two different sensors in Figure \ref{fig:signal} at an OverVoltage (OV, defined as the difference between the applied and the breakdown voltage) of $\sim$ 6 V: it is possible to clearly distinguish signals due to single SPAD events (first peak of about 0.08 a.u.) and signals due to multiple cells events.
In the results reported in this paper the contamination due to DC events was estimated to be $<$~5$\%$ for the single SPAD events, negligible for the multiple SPAD events. 

For the timing analysis a fixed threshold of approximately 20$\%$ of the single SPAD signal amplitude was used for the tested SiPMs, while a CFD (Constant Fraction Discrimination) threshold of 50$\%$ was used for the LGADs used as reference. 

\section{Results}\label{result}

\subsection{Distinguishing between Cherenkov light in  the protection layer and internal processes}
\label{subsec:cherenkov}
As described in the introduction, the first aim was to understand the origin of the excess measured in the CrossTalk (CT) on the SiPM when crossed by a charged particle (henceforth called multi-spad signal) with respect to standard CT-DC (intrinsic CrossTalk measured on Dark Count events) expectation. An explanation based on a Cherenkov effect due to the protection layer implied that the excess should be present with particles impinging the sensor from the front (front here defined as the photon sensitive side), while it should be highly reduced/disappear for particles coming from the back. On the contrary, if the effect is related to processes in the bulk of the sensor structure, then there should be no difference in the response depending on the direction of the impinging particle.

We defined as $\text{F}_\text{n}$  the fraction of events with n firing SPADs w.r.t. the total: 

\begin{equation}
{ \text{F}_\text{n} } = \frac{\text{events with n SPADs firing}}{\text{events with}  \geq 1 \text{ SPADs firing}}
\end{equation}\\ 

\begin{figure}[h!]
        \centering
        \includegraphics[width=0.84\textwidth]{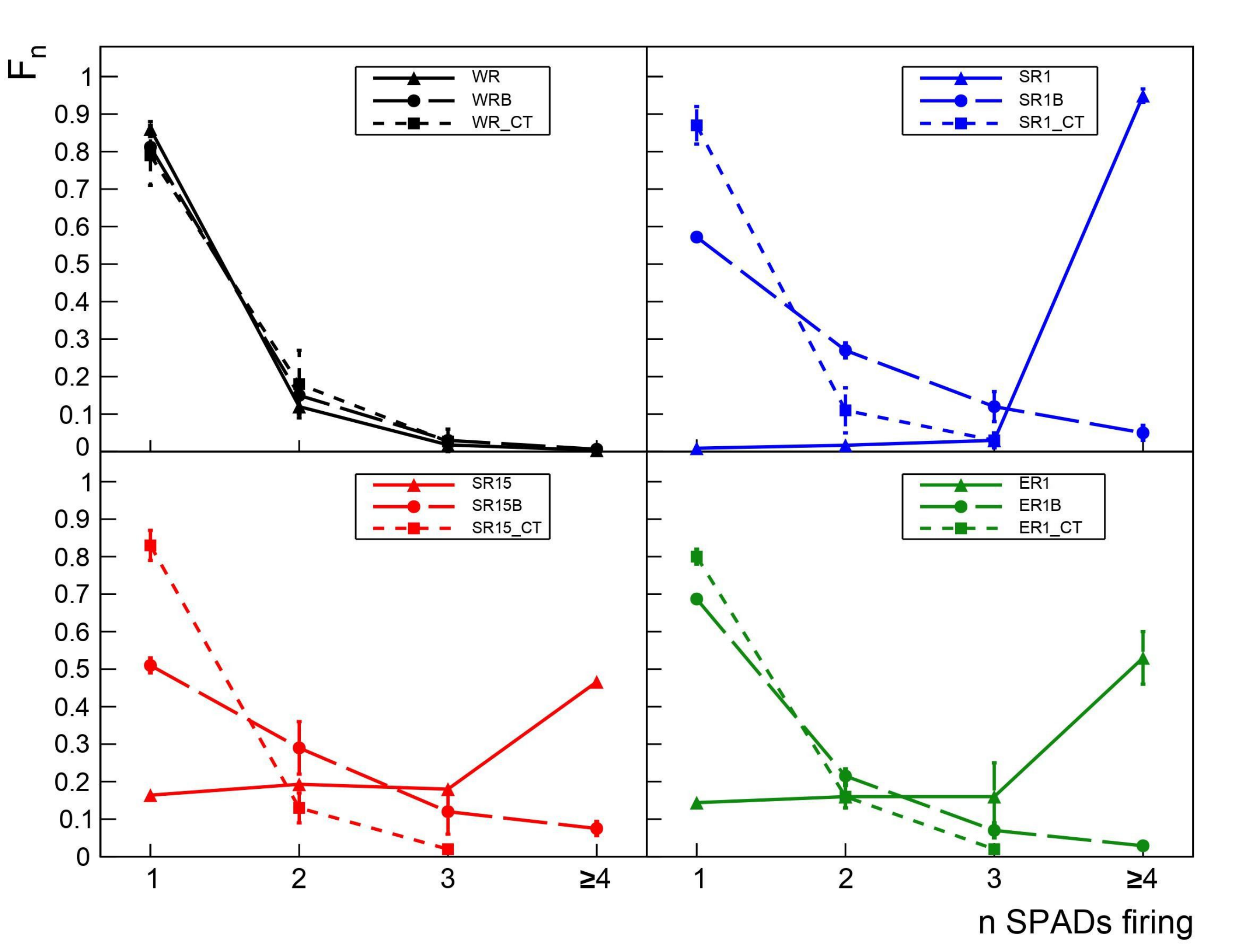}
        \caption{Measured $F_n$  versus the number of fired SPADs for the four sensors and an OV of $\sim$ 6V: a) WR,  b) SR1 , c) SR15 and d)  ER1. Triangular markers indicate measurements with beam from the front, circular markers for beam from the back and square markers for intrinsic CT measured on DC events. 
        }
          \label{fig:firedSPAD}
\end{figure}

The measurement was repeated for different sensors, different voltages and with particles impinging both from front and from back the SiPMs. Due to the amplitude finite scale of the oscilloscope, $n$ can assume values of 1, 2, 3 and $\geq$ 4. Results are reported in Figure \ref{fig:firedSPAD}. 

Figure \ref{fig:firedSPAD} clearly indicates that the effect observed in \cite{SiPM1} is not connected to peculiar processes inside the sensor bulk. In fact, looking at the WR sensor, there is no difference between particle impinging the detector from front or back and, moreover, the measured values are consistent with the CT-DC results indicating mostly one SPAD firing at the passage of the particle.

On the contrary, the other sensors clearly indicate a multi-spad signal when the particle enters the sensor from the front, with large fractions of events with  $\geq$  4 fired SPADs.  For particles coming from the back, it can be also noticed that the distribution of $\text{F}_\text{n}$ is shifted to higher n SPADs values w.r.t. to the case of zero protection layer that could be related to reflections of photons in the protection layer when the particle enters the resin from the back. Notice that the CT-DC of all the SiPMs studied are compatible. \\
These results prove the presence of Cherenkov light during the passage of a charged particle in the SiPM window protection layer (a simple scintillation effect would not cause any difference in a front-back multi-SPAD signal comparison). 

It should be also underlined that the light cone produced, assuming a refraction index of about 1.5 typical of our samples, may reach areas with radius of the order of the 0.5 - 1 mm, making the sensor also able to detect light from a particle passing nearby and not straight across the sensor. Due to the specific feature of a protection layer extending beyond the sensor under test as described in section \ref{sec2}, the alignment of the SiPM with the LGADs may influence the measurement of each sensor in terms of absolute light produced. For this reason the results in Figure \ref{fig:firedSPAD} do not allow a quantitative comparison of the photons produced (or rather, of the number of SPADs firing) among different samples.

\subsection{Timing}
\label{subsec:timing}

\begin{figure}[h!!]
        \centering
        \includegraphics[width=0.7\textwidth]{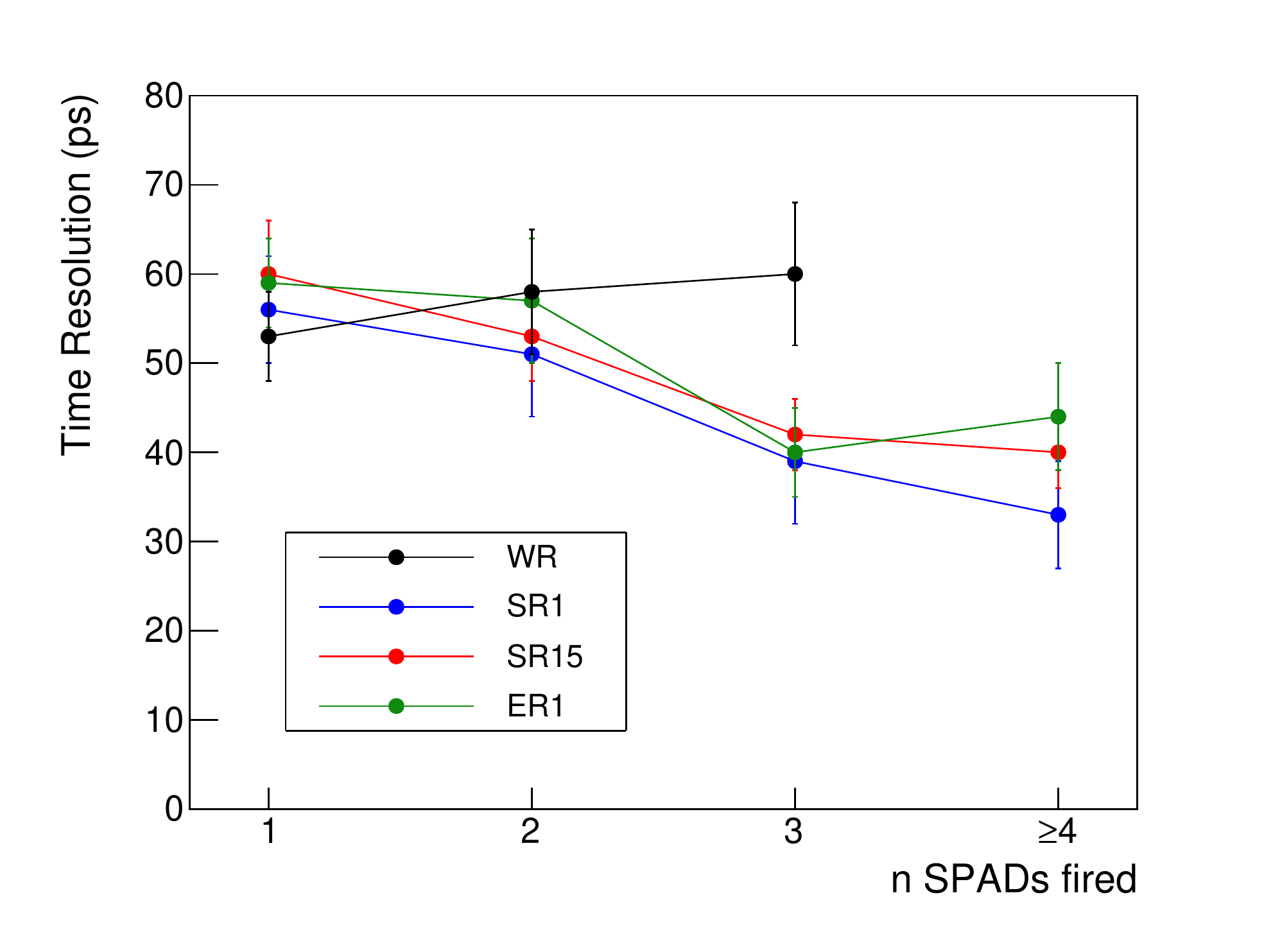}
        \caption{Measured time resolution as a function of the number of SPADs fired for the four sensors described in Section \ref{sec2}. Data correspond to an overvoltage of $\sim$ 6 V. }
          \label{fig:timing}
\end{figure}

In Figure \ref{fig:timing}, the measured time resolution is reported for the four sensors tested as a function of the observed number of SPADs, at an overvoltage of $\sim$ 6 V. The measurement has been performed using one of the reference LGAD, measuring the time difference sensor-LGAD and subtracting the LGAD contribution typically of 27 ps, similarly to \cite{SiPM1}. Note that also here the sigma from a q-Gaussian fit is used.
In all SiPMs with a protection layer the measured value improves with the number of fired SPADs reaching about 30-40 ps and with a dependence from the number of SPADs similar for all the sensors. An overvoltage of $\sim$ 6 V was used, but similar results are obtained at different voltages. For the WR sensor the multi-SPADs events are less probable, as reported in Figure \ref{fig:firedSPAD};  the time resolution when only one SPAD is firing is compatible with what obtained for the other SiPMs. However, with increasing number of SPADs, the time resolution trend seems different from those with resin, not improving with larger number; a possible explanation would be that in the WR sensor, multiple SPAD events are exclusively due to the CT-DC, whereas in the other SiPMs, multiple SPADs are triggered simultaneously by the MIP and associated photons.

The results could benefit from optimized custom FrontEnd electronics.
Moreover, considering the results of the previous section and the fact that the Cherenkov cone is wide and may cover several sensors, summing the outputs of nearby sensors could considerably increase the number of SPADs associated to the traversing particle, improving the signal to noise ratio and the time resolution.Using a single SiPM readout, it could be possible to use the coincident information coming from several SiPMs to improve the timing response.

\section{Conclusions}\label{conclusion}

In this paper a new study of SiPMs response to the passage of a charged particle is reported. 
A resin layer in the entrance window of a SIPM is commonly used as a protective layer. In this paper, the identification of that layer as the source, via Cherenkov light production in it, of the signal registered at the passage of a ionizing charged particle originally observed in \cite{SiPM1} paves the way for moving SiPMs from photosensors to charged particle detectors.

Four different types of SiPMs have been provided by FBK: one without any resin and three with different protection layers, both in thickness and material.
By comparing the number of fired cells with particles impinging the sensor both from the front or the back, the results clearly indicate the production of photons due to the passage of the particle through the protection layer.
The effect produces signals much higher than expected with a benefit also in terms of time resolution, that can reach 30-40 ps in the present setup. More detailed studies will help to better understand the effect of the different protection layer material and thickness; the behavior of sensor with larger coverage should be also investigated.

These results could have important applications in the next generation of RICH counters in combination with a TOF detector or for TOF counters in space. 
Indeed such detector would open the possibility to detect and distinguish at the same time both single photons and charged particles with an excellent time resolution.

\section*{Declarations}
The study was funded by: INFN and FBK.
The authors received research support from institutes as specified in the author list beneath the title. \\

\section*{Data availability}
The datasets generated during and/or analysed during the current study are available from the corresponding author on reasonable request.

\bibliography{sn-bibliography}


\end{document}